\documentclass[sigconf,natbib=true,anonymous=false]{acmart}%

\usepackage{amsmath,amsfonts}
\usepackage{amsthm}
\usepackage{graphicx}
\usepackage{textcomp}
\usepackage{balance}
\usepackage{color,xcolor}
\usepackage{multirow}
\usepackage{subfigure}
\usepackage{array}
\usepackage{enumitem}
\usepackage{epstopdf}
\usepackage{booktabs}
\usepackage{threeparttable}
\usepackage{bm}
\usepackage{hyperref}

\newtheoremstyle{mydefn}
{}{}
{\it}  
{0pt}       
{\bfseries} 
{:}
{.5em}
{}          
\theoremstyle{mydefn}

\newtheorem{definition}{Definition}

\newtheoremstyle{myexample}
{}{}
{}  
{0pt}       
{\bfseries} 
{:}
{.5em}
{}          
\theoremstyle{myexample}

\newtheorem{example}{Example}

\renewcommand{\paragraph}[1]{\vspace{0.5em}\noindent\textbf{#1.}}
\renewcommand{\subparagraph}[1]{\vspace{0.5em}\noindent\textit{\underline{#1.}}}

\newcommand{\floor}[1]{\lfloor #1 \rfloor}

\newcommand{\norm}[1]{\Vert #1 \Vert}
\newcommand{\num}[1]{\vert #1 \vert}

\makeatletter
\newif\if@restonecol
\makeatother

\usepackage[linesnumbered,ruled,vlined]{algorithm2e}
\usepackage{algpseudocode}

\SetKwComment{Comment}{$\triangleright$\ }{}

\AtBeginDocument{%
  \providecommand\BibTeX{{%
    \normalfont B\kern-0.5em{\scshape i\kern-0.25em b}\kern-0.8em\TeX}}}
    
\setcopyright{acmcopyright}
\copyrightyear{2021}
\acmYear{2021}
\acmDOI{10.1145/1122445.1122456}

\acmConference[Conf '21]{Conf '21: Proceedings of 2021 ACM XXX Conference}{June, 2021}{Online}
\acmBooktitle{Conf '21: Proceedings of 2021 ACM XXX Conference, June, 2021, Online}
\acmPrice{15.00}
\acmISBN{978-1-4503-XXXX-X/18/06}

\begin{document}

\title{A Generic Distributed Clustering Framework for Massive Data}

\author{Pingyi Luo}
\affiliation{
	\institution{School of Computing}
	\institution{National University of Singapore}
	\country{Singapore} 
}
\email{pingyi@u.nus.edu}

\author{Qiang Huang}
\affiliation{
	\institution{School of Computing}  	
  	\institution{National University of Singapore}
  	\country{Singapore} 
}
\email{huangq@comp.nus.edu.sg}

\author{Anthony K. H. Tung}
\affiliation{
	\institution{School of Computing}
  	\institution{National University of Singapore}
  	\country{Singapore}
}
\email{atung@comp.nus.edu.sg}

\renewcommand{\shortauthors}{Pingyi Luo, et al.}

\begin{abstract}
In this paper, we introduce a novel Generic distributEd clustEring frameworK (GEEK) beyond $k$-means clustering to process massive amounts of data. To deal with different data types, GEEK first converts data in the original feature space into a unified format of buckets; then, we design a new Seeding method based on simILar bucKets (SILK) to determine initial seeds. Compared with state-of-the-art seeding methods such as $k$-means++ and its variants, SILK can automatically identify the number of initial seeds based on the closeness of shared data objects in similar buckets instead of pre-specifying $k$. Thus, its time complexity is independent of $k$. With these well-selected initial seeds, GEEK only needs a one-pass data assignment to get the final clusters. We implement GEEK on a distributed CPU-GPU platform for large-scale clustering. We evaluate the performance of GEEK over five large-scale real-life datasets and show that GEEK can deal with massive data of different types and is comparable to (or even better than) many state-of-the-art customized GPU-based methods, especially in large $k$ values. 
%
\end{abstract}

%


\maketitle

\section{Introduction}
\label{sect:intro}
Over the last decade, data sets have emerged on an unprecedented scale across different scientific disciplines \cite{bachem2017distributed}. 
The data types in different scenarios may be quite different, i.e., dense or sparse. Moreover, the heterogeneous data with the co-existence of categorical and numeric attributes become very popular. 
This development has created a new demand for distributed data mining and machine learning methods to deal with massive amounts of data of different data types. 
This paper considers the generic distributed clustering beyond $k$-means, which is useful in two ways.

First, as a method to find clusters, many partitioning based methods like $k$-means are linear in complexity in terms of both cardinality (i.e., $n$) and dimensionality (i.e., $d$) of the data. Their results can also be easily interpreted by looking at the representative of each cluster that is found. Comparatively, other methods like hierarchical \cite{zhang1996birch, guha1998cure}, density-based \cite{ester1996density, ankerst1999optics, gan2015dbscan} and grid-based clustering \cite{wang1997sting, agrawal1998automatic} are typically more expensive especially in high-dimensional spaces \cite{han2001spatial,gan2015dbscan}. Interpretability wise, while hierarchical methods can form dendrograms to facilitate the understanding of these clusters, understanding the clusters that are discovered through density or grid based methods is much harder.

Second, using partitioning methods such as $k$-means to preprocess the data into a large number of microclusters (i.e., $k$ sufficiently high but $k \ll n$) \cite{jin2001mining}, other kinds of clustering methods can speed up substantially. This can be done in two ways. First, for hierarchical methods, one can directly use the microclusters as the base objects to form hierarchical clusters instead of using the actual data points which are much higher in numbers than the microclusters. Second, one can utilize the microclusters to build indexes to support the efficient execution of neighborhood queries in density or grid-based methods \cite{johnson2019billion}. Both of them rely on the efficiency (i.e., can handle large values of $n$ and $k$) and effectiveness (i.e., high-quality microclusters with low diameters) of the partitioning methods. In view of the usefulness of partitioning methods such as $k$-means, we will look at their speedup in this paper.

A popular version of $k$-means is the \emph{Lloyd's algorithm} \cite{lloyd1982least}, which is conducted repeatedly by \emph{assigning} data objects to their closest central vectors (e.g., centroids) and then \emph{updating} the $k$ central vectors by their assigned members, such that the total distances between data objects and their closest central vectors are minimized. Various methods and optimizations are proposed based on this algorithm to deal with the problems of initial seeds selection \cite{arthur2007k, bahmani2012scalable, bachem2016fast, bachem2017distributed, cohen2020fast}, the large values of $d$ and $k$ \cite{elkan2003using, hamerly2010making, ding2015yinyang, newling2016nested, curtin2017dual}, and the acceleration for massive data (i.e., $n$) with modern hardware \cite{cao2006scalable,li2013speeding,li2018large}. However, most of them are highly optimized for a specific distance or similarity measure, e.g., Euclidean distance or its variants. To the best of our knowledge, there is no generic $k$-means variant that can deal with different types of data.

On the other hand, Locality-Sensitive Hashing (LSH) \cite{indyk1998approximate, har2012approximate} and its variants \cite{broder1998min, charikar2002similarity, datar2004locality, andoni2006near, shrivastava2014asymmetric, andoni2015practical,  huang2018accurate, lei2019sublinear, lu2020vhp, zheng2020pm, lei2020locality} are often used for clustering.
The most typical usage of LSH is to accelerate the Nearest Neighbor Search (NNS) during clustering \cite{koga2007fast, li2014efficient, cohen2020fast}. For example, Cohen-Addad et al. leveraged LSH to shortlist the candidates to compare with the centroids, so that the expensive distance computations can be vastly relieved \cite{cohen2020fast}. 
Another common scenario is to approximate the distance with LSH to avoid actual computation \cite{zamora2016hashing, de2016hash}. In \cite{zamora2016hashing}, they used MinHash \cite{broder1998min} and SimHash \cite{charikar2002similarity} to approximate the distance for high-dimensional data (e.g., text) and applied a bisection method to the LSH similarity matrix for clustering.
For the large-scale, distributed setting, LSH is also applied for data partitioning to reduce communication cost, e.g., \cite{zhang2016efficient, bhaskara2018distributed} exploited LSH to distribute close data on the same machine in a distributed environment and find clusters locally; then, they aggregated local results to approximate the global clusters.

Most of these methods, however, only use LSH as a basic tool for speedup or data partitioning. They focus on the data objects in a single bucket but do not consider \emph{the connections between buckets}. 
Intuitively, similar buckets often share many common data objects under many LSH functions. These shared data are close to each other, which are the ideal candidates to determine initial seeds for clustering. 
Moreover, there exist different kinds of LSH families for different types of data. With different LSH families, different types of data can be converted into a unified format of buckets. 

\paragraph{Our Contributions}
Based on the above motivations, we introduce a novel \underline{G}eneric distribut\underline{E}d clust\underline{E}ring framewor\underline{K} (GEEK) to deal with large-scale data of different types. 
We first leverage different LSH families to convert different types of data into a unified format of buckets. Then, we design a new \underline{S}eeding algorithm based on sim\underline{IL}ar buc\underline{K}ets (SILK) to generate initial seeds. Specifically, we adapt MinHash \cite{broder1997resemblance, broder1998min}, which is a popular LSH scheme for NNS between sets, to retrieve similar buckets. Then, we use the shared data objects from similar buckets to determine the initial seeds. 
As distinct from state-of-the-art seeding methods such as $k$-means++ and its variants \cite{arthur2007k, bahmani2012scalable, bachem2017distributed, cohen2020fast}, SILK can automatically identify initial seeds without pre-specifying $k$, and its time complexity is independent of $k$. 
With these well-selected initial seeds, we get the final clusters through a one-pass data assignment. 
We exploit the parallelism with multiple computing nodes and GPUs and implement GEEK on a distributed CPU-GPU platform for large-scale clustering. 
Extensive experiments over five massive real-life datasets validate the efficiency and effectiveness of GEEK.

\paragraph{Organization}
The roadmap of this paper is as follows. Section \ref{sect:pre} reviews some preliminary knowledge. We introduce GEEK and its distributed implementation in Section \ref{sect:method}. Experimental results are reported and analysed in Section \ref{sect:expt}. Section \ref{sect:related} surveys the related work. Finally, we conclude our work in Section \ref{sect:conclusion}.

\section{Preliminaries}
\label{sect:pre}

\subsection{Problem Settings}
\label{sect:pre:problem}
In this paper, we focus on both dense and sparse data, where dense data consists of data objects whose values in most feature dimensions (or attributes) are non-zero, whereas the sparse data mainly contains zeros. 
The dense data can be further divided into the \emph{homogeneous} dense data (with single numeric or categorical attribute) and the \emph{heterogeneous} dense data (with the co-existence of numeric and categorical attributes). 
The sparse data is often homogeneous; otherwise, we can encode the data into a single attribute (e.g., one-hot encoding) due to the ultra-high dimensions. 

Suppose data objects are represented as vectors in a $d$-dimensional space $\mathbb{R}^d$. Let $Dist(\bm{x},\bm{y})$ be the distance between any $\bm{x},\bm{y} \in \mathbb{R}^d$.
\begin{definition}[$k$-means clustering]
\label{def:clustering}
Given a data set $\mathcal{D}$ of $n$ data objects, i.e., $\mathcal{D} = \{\bm{x_1},\cdots,\bm{x_n}\} \subset \mathbb{R}^d$, $k$-means clustering aims to find $k$ central vectors $\mathcal{C} = \{\bm{c_1},\cdots,\bm{c_k}\} \subset \mathbb{R}^d$ such that the total distance $Dist(\mathcal{D},\mathcal{C})$ between all of the data objects and their closest central vectors is minimized, where $Dist(\mathcal{D},\mathcal{C}) = \sum_{i=1}^{n} Dist(\bm{x_i},\bm{c_{{ID}_i}})$, and ${ID}_i$ is the index of the closest central vector of $\bm{x_i}$, i.e.,
\begin{equation}
\label{eqn:closest-center}
{ID}_i = {\arg \min}_{j \in \{1,\cdots,k\}} Dist(\bm{x_i},\bm{c_j}). 
\end{equation}
\end{definition}

$Dist(\bm{x},\bm{y})$ can be widespread distance functions, such as Euclidean distance, Manhattan distance, Hamming distance, etc. Given a similarity $Sim(\bm{x},\bm{y}) \in [0,1]$, e.g., Jaccard similarity, we consider its corresponding distance $Dist(\bm{x},\bm{y}) = 1 - Sim(\bm{x},\bm{y})$ in Equation \ref{eqn:closest-center}. 
Notice that we do \emph{not} claim that GEEK can support every distance and similarity measure -- just the distances or similarity measures if and only if there exists the corresponding LSH family.

\subsection{Locality-Sensitive Hashing}
\label{sect:pre:lsh}
\begin{definition}[LSH family \cite{indyk1998approximate}]
\label{def:lsh_family}
Given a search radius $R > 0$ and an approximation ratio $c$, a family of hash functions $\mathcal{H}$ is $(R,cR,p_1,p_2)$-sensitive for $Dist(\cdot,\cdot)$ if, for any $\bm{x},\bm{y} \in \mathbb{R}^d$, the hash function $h$ drawn uniformly from $\mathcal{H}$ satisfies the following conditions:
\begin{itemize}
\item If $Dist(\bm{x},\bm{y}) \leq R$, then $\Pr_{h \in \mathcal{H}} [h(\bm{x}) = h(\bm{y})] \geq p_1$;
\item If $Dist(\bm{x},\bm{y}) \geq cR$, then $\Pr_{h \in \mathcal{H}} [h(\bm{x}) = h(\bm{y})] \leq p_2$;
\item $p_1 > p_2$ and $c>1$.
\end{itemize}
\end{definition}

Next, we review two famous LSH families, MinHash \cite{broder1997resemblance, broder1998min} and Query-Aware LSH (QALSH) \cite{huang2015query, huang2017query}, which will be used later.

\paragraph{MinHash} 
MinHash is designed for Jaccard similarity $J(A, B) = |A \cap B|/|A \cup B|$ between any two sets (or objects) $A,B \in \mathbb{U}$. The MinHash function $h_\pi(\cdot)$ is defined as follows: 
\begin{equation}
\label{eqn:lsh-minhash}
h_\pi(A) = \min_{a \in A} \pi(a),
\end{equation}
where $\pi$ is a random permutation from the ground universe $\mathbb{U}$ and $h_\pi(A)$ is the minimal member of $A$ with respect to $\pi$.

MinHash uses the static $(K,L)$-bucketing framework \cite{indyk1998approximate} to deal with the NNS query. It concatenates $K$ i.i.d. LSH functions to form a signature, i.e., $G(\cdot) = (h_{\pi_1}(\cdot),\cdots,h_{\pi_K}(\cdot))$. Since $G(\cdot)$ will reduce $p(s)$ between close objects, MinHash builds $L$ such hash tables to achieve a theoretical guarantee. MinHash looks up the $L$ buckets based on the signatures of the query to find the candidates.

\paragraph{QALSH} 
QALSH is developed for Euclidean distance $\norm{\bm{x}-\bm{y}} = \sqrt{\sum_{j=1}^d (x_j - y_j)^2}$, where $\bm{x}=(x_1,\cdots,x_d)$ and $\bm{y}=(y_1,\cdots,y_d)$. The QALSH function $h_{\bm{a}}(\cdot)$ is defined as follows: 
\begin{equation}
\label{eqn:lsh-qalsh}
h_{\bm{a}}(\bm{x}) = \bm{a} \cdot \bm{x},
\end{equation}
where $\bm{a}$ is a random vector with each entry $a_i$ chosen i.i.d. from the standard normal distribution, i.e., $a_i \sim \mathcal{N}(0,1)$. 

QALSH adopts the dynamic collision counting framework \cite{gan2012locality} for the NNS. It draws $m$ i.i.d. LSH functions to build $m$ hash tables, where each stores the sorted LSH values only. Let $w$ be the bucket width. Given a query $\bm{q}$, the interval $[h_{\bm{a}}(\bm{q})-\frac{w}{2},h_{\bm{a}}(\bm{q})+\frac{w}{2}]$ is regarded as the anchor bucket of $\bm{q}$. The bucket partitioning waits until $\bm{q}$ arrives. We select an object as a candidate of $\bm{q}$ if they collide more than $l$ times, where $l$ is a pre-specified collision threshold.

\section{GEEK}
\label{sect:method}
\begin{figure*}[htb]
\centering
\includegraphics[width=0.98\textwidth]{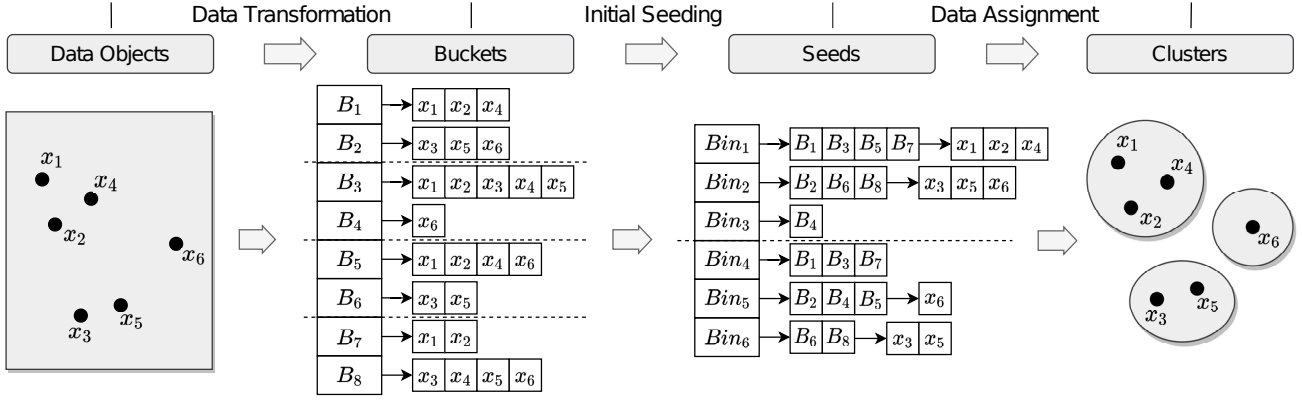}
\vspace{-1.0em}
\caption{An overview of GEEK}
\label{fig:example}
\vspace{-0.75em}
\end{figure*}

We now present GEEK for generic distributed clustering. GEEK consists of three phases: data transformation, initial seeding, and data assignment. An overview of GEEK is shown in Figure \ref{fig:example}.

\subsection{Data Transformation}
\label{sect:method:trans}
In this phase, we convert different types of data into a unified format of buckets. We illustrate one specific distance for each data type for demonstration purposes.

\begin{algorithm}[t]
\caption{Homogeneous Dense Data Transformation}
\label{alg:transform-homo}
\KwIn{$\mathcal{D} = \{\bm{x_1},\cdots,\bm{x_n}\}$, $m$, $t$ for bucket partitioning;}
\KwOut{A collection of buckets $\mathcal{B}$;}
$\mathcal{B} = \emptyset$\; 
Draw $m$ QALSH functions $\{h_{\bm{a_1}},\cdots,h_{\bm{a_m}}\}$\;
\For {$i$ from $1$ to $m$} {
	$T = \emptyset$\;  	
  	\For {each $\bm{x_j} \in \mathcal{D}$} {
    	$T.push(h_{\bm{a_i}}(\bm{x_j}))$\;
  	}
  	Sort $T$ in ascending order of $h_{\bm{a_i}}(\bm{x_j})$\;
  	Partition $T$ into $t$ buckets $\{B_1,\cdots,B_t\}$\;
  	$\mathcal{B} = \mathcal{B} \cup \{B_1,\cdots,B_t\}$\;
}
\Return $\mathcal{B}$\;
\end{algorithm}

\paragraph{Homogeneous Dense Data}
We consider high dimensional points (or vectors) for Euclidean distance. We first draw $m$ i.i.d. QALSH functions and build $m$ hash tables. Let $t$ be a threshold for bucket partitioning. Then, we sort the $m$ hash tables in ascending order of the LSH values and evenly partition the data IDs in each hash table into $t$ buckets. Let $\mathcal{B}$ be a collection of buckets generated from the $m$ hash tables. The pseudo-code is depicted in Algorithm \ref{alg:transform-homo}.

\begin{example}
\label{exp:transformation}
We now use Figure \ref{fig:example} as an illustration. Suppose $\mathcal{D} = \{\bm{x_1},\bm{x_2},\cdots,\bm{x_6}\}$ and there exist $3$ clusters, i.e., $\{\bm{x_1},\bm{x_2},\bm{x_4}\}$, $\{\bm{x_3},\bm{x_5}\}$, and $\{\bm{x_6}\}$. We use $m=4$ QALSH functions $\{h_{\bm{a_1}},\cdots,h_{\bm{a_4}}\}$ to transform all $\bm{x_j} \in \mathcal{D}$ into buckets. Let $t=2$. For the hash table $T_1$, we compute $h_{\bm{a_1}}(\bm{x_j})$ for each $\bm{x_j}$ and partition all of them into two buckets, i.e., $B_1 = \{\bm{x_1},\bm{x_2},\bm{x_4}\}$ and $B_2 = \{\bm{x_3},\bm{x_5},\bm{x_6}\}$. We repeat this procedure for $T_2$, $T_3$, and $T_4$ and get $\mathcal{B} = \{B_1,B_2,\cdots,B_8\}$.
\hfill $\triangle$ \par 
\end{example}

\vspace{-0.5em}
\subparagraph{Remarks}
Notice that we do not use the popular LSH function $h_{\bm{a},b}(\bm{x}) = \floor{\tfrac{\bm{a}\cdot\bm{x}+b}{w}}$ \cite{datar2004locality} for bucket partitioning, because it contains a parameter named bucket width $w$ that is hard to tune for different datasets if we want to partition data into buckets in a suitable granularity. 
Yet, we do not simply follow QALSH and use $[h_{\bm{a}}(\bm{x})-\frac{w}{2},h_{\bm{a}}(\bm{x})+\frac{w}{2}]$ as the bucket of each data object $\bm{x}$, because this way is customized for the query, which is not suitable for data objects. Instead of using $w$, we introduce a new parameter $t$ to evenly partition each hash table into $t$ parts as buckets. This new way is simple and easy to control the granularity of buckets, and it can keep the proximity of most data objects in the same buckets.

\vspace{-0.5em}
\paragraph{Heterogeneous Dense Data}
The categorical attributes are discrete values, which is a natural division for buckets. However, the numeric attributes are continuous values, which cannot be divided into buckets directly. Thus, we first follow the homogeneous dense data transformation to discretize the numeric attributes into categorical attributes. 
We then follow MinHash and leverage the static $(K,L)$-bucketing framework to transform the categorical attributes into buckets. The pseudo-code is shown in Algorithm \ref{alg:transform-hetero}.

\begin{algorithm}[t]
\caption{Heterogeneous Dense Data Transformation}
\label{alg:transform-hetero}
\KwIn{$\mathcal{D} = \{\bm{x_1},\cdots,\bm{x_n}\}$, $K$, $L$ for bucket partitioning;}
\KwOut{A collection of buckets $\mathcal{B}$;}
Transform numeric attributes into categorical ones\;
$\mathcal{B} = \emptyset$\;
\For {$i$ from $1$ to $L$} {
	Draw $K$ MinHash functions $\{h_{\pi_1},\cdots,h_{\pi_K}\}$\;	
	$T = \emptyset$\;	
  	\For {each $\bm{x_j} \in \mathcal{D}$} {
    	$G(\bm{x_j}) = (h_{\pi_{1}}(\bm{x_j}),\cdots,h_{\pi_{K}}(\bm{x_j}))$\;
    	$T[G(\bm{x_j})].push(j)$\;
  	}
  	$\mathcal{B} = \mathcal{B} \cup T$\;
}
\Return $\mathcal{B}$\;
\end{algorithm}

\paragraph{Sparse Data}
We consider data objects represented as sets for Jaccard similarity. 
Since sparse datasets are often ultra-high dimensional, e.g., million-scale dimensions, we first follow \cite{wang2018randomized} and use DOPH \cite{shrivastava2014densifying} to reduce the dimensionality to a moderate level, e.g., hundreds of dimensions.\footnote{In our experiments, for the sparse dataset URL, we first use DOPH to reduce the data dimension to 400.} We choose DOPH for dimensionality reduction because with DOPH, the distance among each pair of data after dimension reduction can still be approximately preserved with high probability \cite{wang2018randomized}. 
Then, we follow MinHash and use the static $(K,L)$-bucketing framework to transform data objects into buckets. The pseudo-code is displayed in Algorithm \ref{alg:transform-sparse}.

\begin{algorithm}[t]
\caption{Sparse Data Transformation}
\label{alg:transform-sparse}
\KwIn{$\mathcal{D} = \{\bm{x_1},\cdots,\bm{x_n}\}$, $K$, $L$ for bucket partitioning;}
\KwOut{A collection of buckets $\mathcal{B}$;}
Reduce $\mathcal{D}$ to moderate dimensions using DOPH \cite{shrivastava2014densifying}\;
$\mathcal{B} = \emptyset$\;
\For {$i$ from $1$ to $L$} {
	Draw $K$ MinHash functions $\{h_{\pi_1},\cdots,h_{\pi_K}\}$\;	
	$T = \emptyset$\;
  	\For {each $\bm{x_j} \in \mathcal{D}$} {
    	$G(\bm{x_j}) = (h_{\pi_{1}}(\bm{x_j}),\cdots,h_{\pi_{K}}(\bm{x_j}))$\;
    	$T[G(\bm{x_j})].push(j)$\;
  	}
  	$\mathcal{B} = \mathcal{B} \cup T$\;
}
\Return $\mathcal{B}$\;
\end{algorithm}

\subsection{Initial Seeding}
\label{sect:method:seeding}
We now introduce SILK to generate initial seeds based on similar buckets. 
The idea of SILK is first to adapt MinHash to group similar buckets together, and then we identify the shared data objects from similar buckets to determine the initial seeds.

We first adopt the static $(K,L)$-bucketing framework to transform the collection of buckets $\mathcal{B}$ into bins.\footnote{To distinguish the buckets from data transformation, we call the buckets generated from SILK as bins. Each bin is a set of bucket IDs.} 
For each bin ${Bin}_j$, if it contains at most one bucket, i.e., $\num{{Bin}_j} \leq 1$, we ignore it because this bucket is probably isolated from others; otherwise, we identify a group of shared data objects (i.e., $C_{shared}$) from ${Bin}_j$ by \emph{majority voting}, i.e., we select the data objects that appear in more than half number of the buckets in ${Bin}_j$. 
With majority voting, we can identify the frequent shared data objects from similar buckets. Intuitively, the higher the frequency, the closer the distance to other shared data objects. 

After identifying a collection of $C_{shared}$'s, we need to decide which $C_{shared}$ can be used to determine initial seeds. Intuitively, the larger the size of $C_{shared}$, the higher the probability that there exists a cluster among the data objects in $C_{shared}$. 
Let $\delta$ be a seeding threshold and $\mathcal{C}$ be a collection of initial seeds. Based on this intuition, if $\num{C_{shared}} \geq \delta$, we will add it into $\mathcal{C}$. 

However, when we repeat the above procedure to identify the initial seeds $\mathcal{C}$ for the $L$ hash tables, some very similar or even identical $C_{shared}$'s may be added into $\mathcal{C}$ multiple times. 
Fortunately, $\mathcal{C}$ can also be regarded as a set of buckets. We can use $\mathcal{C}$ as input and repeat the above procedure again to remove the near-duplicates. The pseudo-code is depicted in Algorithm \ref{alg:silk}.

\begin{example}
\label{exp:seeding}
We now use Figure \ref{fig:example} again to illustrate SILK. We use $\mathcal{B} = \{B_1,B_2,\cdots,B_8\}$ as input. Suppose $L=2$ and $\delta = 1$. We first partition similar buckets into the same bins, e.g., for hash table $T_1$, we get ${Bin}_1 = \{B_1,B_3,B_5,B_7\}$, ${Bin}_2 = \{B_2,B_6,B_8\}$, and ${Bin}_3 = \{B_4\}$. 
Then, we identify $C_{shared}$ from each ${Bin}_j$. Considering ${Bin}_1$, since $\bm{x_1},\bm{x_2},\bm{x_4}$ appear in more than half number of $B_1,B_3,B_5,$ and $B_7$, we add $\bm{x_1},\bm{x_2},\bm{x_4}$ into $C_{shared}$; while $\bm{x_3},\bm{x_5},\bm{x_6}$ only appear once, we ignore them. Since $\num{C_{shared}} = 3 \geq \delta = 1$, we add $C_{shared} = \{\bm{x_1},\bm{x_2},\bm{x_4}\}$ into $\mathcal{C}$. We ignore ${Bin}_3$ since $\num{{Bin}_3} \leq 1$. 
After identifying $C_{shared}$'s, some very similar or even identical $C_{shared}$'s are added into $\mathcal{C}$, e.g., we also add $C_{shared} = \{\bm{x_1},\bm{x_2},\bm{x_4}\}$ from ${Bin}_4$ into $\mathcal{C}$. Thus, we repeat this procedure again with $L=1$ to deduplicate $C_{shared}$'s. Finally, we get four $C_{shared}$'s in $\mathcal{C}$, i.e., $\{\bm{x_1},\bm{x_2},\bm{x_4}\}$, $\{\bm{x_3},\bm{x_5},\bm{x_6}\}$, $\{\bm{x_6}\}$, and $\{\bm{x_3},\bm{x_5}\}$.
\hfill $\triangle$ \par 
\end{example}

\paragraph{Remarks}
The state-of-the-art seeding methods, such as $k$-means++ and its variant {\cite{arthur2007k, bahmani2012scalable, bachem2017distributed, cohen2020fast}, require setting up $k$ in advance. Moreover, their time complexities are highly sensitive to $k$, which limit their scalability for large $k$ values. In contrast, as will be discussed in Section \ref{sect:impl:complexity}, even though SILK needs to tune a few parameters, i.e., $K$, $L$, and $\delta$, its time complexity is independent of $k$, and it can efficiently generate a large number of initial seeds. 

\begin{algorithm}[t]
\caption{SILK}
\label{alg:silk}
\KwIn{$\mathcal{B}$, a seeding threshold $\delta$, and $K,L$ for partitioning;}
\KwOut{A collection of initial seeds $\mathcal{C}$;}
$\mathcal{C} \gets \emptyset$ \;
\For {$i$ from $1$ to $L$} {
  	Draw $K$ MinHash functions $\{h_{\pi_1},\cdots,h_{\pi_K}\}$\;
	$T \gets \emptyset$\;
  	\For {each bucket $B_j \in \mathcal{B}$} {
    	$G(B_j) = (h_{\pi_1}(B_j),\cdots,h_{\pi_K}(B_j))$\;
    	$T[G(B_j)].push(B_j)$\; 
  	}
  	\For {each ${Bin}_j \in T$} {
		\textbf{if} $\num{{Bin}_j} \leq 1$ \textbf{then} \textbf{continue}\;	
    	Find $C_{shared}$ from the buckets in ${Bin}_j$\;
    	\textbf{if} $\num{C_{shared}} \geq \delta$ \textbf{then} $\mathcal{C} = \mathcal{C} \cup C_{shared}$\;
	}
}
Remove the near duplications of $\mathcal{C}$\;
\Return $\mathcal{C}$\;
\end{algorithm}

\begin{figure*}[htb]
\centering
\includegraphics[width=1.0\textwidth]{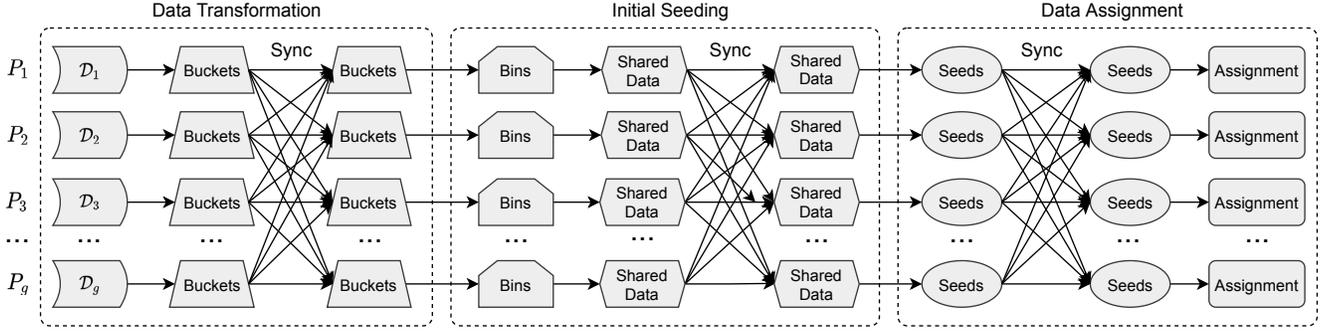}%
\vspace{-0.5em}
\caption{The system architecture of GEEK}
\label{fig:architecture}
\vspace{-0.5em}
\end{figure*}

\subsection{Data Assignment}
\label{sect:method:assign}
In this phase, we first compute the central vectors from the initial seeds $\mathcal{C}$; then, we assign each data object to its closest central vector once to achieve the final clusters. With the efficiency of SILK, we often generate more initial seeds to get more options to detect small clusters. Thus, one-pass data assignment is sufficient. 

\begin{example}
\label{exp:assignment}
From Figure \ref{fig:example}, if we just build a single hash table $T_1$, GEEK generates two initial seeds from $C_{shared_1} = \{\bm{x_1},\bm{x_2},\bm{x_4}\}$ and $C_{shared_2} = \{\bm{x_3},\bm{x_5},\bm{x_6}\}$ only. The small cluster $\{\bm{x_6}\}$ will be combined together with the cluster $\{\bm{x_3},\bm{x_5}\}$ and cannot be distinguished. Thus, we often build more hash tables to get more initial seeds. As shown in Figure \ref{fig:example}, with one more $T_2$ to get more seeds from $C_{shared_3} = \{\bm{x_6}\}$ and $C_{shared_4} = \{\bm{x_3},\bm{x_5}\}$, we can easily identify $\{\bm{x_6}\}$ and avoid multiple assign-update iterations. 
\hfill $\triangle$ \par 
\end{example}

There are several ways to get more initial seeds. In Example \ref{exp:assignment}, we use more hash tables in SILK to generate more seeds. Meanwhile, we can also generate more buckets from data transformation or decrease the seeding threshold $\delta$ to get more seeds.

\paragraph{Central Vector Representation}
For the homogeneous dense data such as the high-dimensional points, we use \emph{centroid} as the central vector. 
For the heterogeneous dense data, it is hard to define a central vector for clustering with its original representation. Thus, we use the unified categorical attributes for data assignment and use \emph{mode} to represent the central vector.
For the sparse data such as sets, we also use \emph{mode} as the central vector. Moreover, since the original dimension is often very large, we use the dimension after DOPH to represent data objects to reduce the computational cost.

\subsection{Implementation}
\label{sect:method:impl}
Next, we introduce how we design and implement GEEK on a distributed CPU-GPU platform.

\paragraph{System Architecture}
Suppose we have $g$ GPUs in the system across $N$ computing nodes. Each node can have different number of GPUs, i.e., $g_i$ for node $i$ and $\sum_{i=1}^N g_i = g$. We adopt the Message Passing Interface (MPI) for the communication of different computing nodes. We launch $g$ processes with MPI. Each process $P_j$ corresponds to a single GPU for parallel computation, where $1 \leq j \leq g$. The system architecture is depicted in Figure \ref{fig:architecture}.

\subparagraph{Data Transformation}
We first instruct GPU threads to compute the LSH values for the data objects that each process $P_j$ contains. Since each $P_j$ only has a part of the data, the buckets in different processes are incomplete. We need to conduct a \emph{bucket synchronization} among different processes to rearrange the buckets so that (1) data objects with the identical LSH values are put into the same buckets; (2) each $P_j$ contains different hash tables. This procedure aims to deal with a \emph{load balance} problem, which will be discussed later.

\subparagraph{Initial Seeding}
After the bucket synchronization, we instruct the GPU threads to compute bins for the buckets that each process $P_j$ contains. Similar to the case of data transformation, the bins in different processes are incomplete. However, to reduce the \emph{communication cost}, we do not rearrange the bins. Instead, we determine the shared data objects $C_{shared}$ for each $P_j$ from its local bins and synchronize the $C_{shared}$'s among different processes. We will introduce the details of this procedure after the load balance issue.

\subparagraph{Data Assignment}
Since different data are located in different nodes, we cannot directly compute the central vectors (e.g., centroids). To reduce the communication cost, we first instruct the GPU threads to compute the local centroids and count their local size for each process $P_j$. Then, we broadcast the local centroids and their sizes to get the global centroids. At last, each $P_j$ performs a one-pass data assignment and gets the final clusters.  

\paragraph{Load Balance}
Load balance is an essential issue for distributed clustering. It consists of two aspects: original data load balance and intermediate data load balance. For the original data, we evenly split the dataset into $g$ parts so that each process has an equal portion of the dataset. Thus, the workload regarding the original data such as data transformation and data assignment are balanced.

For the intermediate data, the primary bottleneck is how to synchronize the buckets to keep the workload balance for SILK during the phase of initial seeding. A direct way is to assign the number of buckets to $g$ processes as even as possible. 
However, different buckets might have a different number of data IDs. Notice that we only evenly partition data objects into buckets for the homogeneous dense data. 
Some processes with small buckets will be idle, while other processes with large buckets may be hectic. In this way, the workload for SILK is imbalanced. 

As will be discussed in Section \ref{sect:impl:complexity}, the time complexity of SILK is determined by the total number of data IDs among the buckets, so we evenly assign the number of hash tables to $g$ processes. 
The insight is that each hash table may have a different number of buckets, but it contains the same number of data IDs. Furthermore, we can control the number of hash tables (e.g., $L$) divisible by $g$. Thus, the intermediate workload for SILK can be balanced by this way of bucket synchronization. 

\subparagraph{Multi-Loading Strategy}
The GPU memory might not be sufficient for the billion-scale or even larger-scale data to keep a single hash table. To fix this problem, we implement a \emph{multi-loading} strategy as follows. Each process runs SILK multiple times; each time, we load as many buckets as possible to fit the GPU memory. Since the GPU memory of each process can be identical, the workload of each process will not be much affected by this multi-loading strategy.

\begin{table*}[t]
\centering
\caption{The time and space complexities of GEEK for the Homo, Hetero, and Sparse data}
\vspace{-0.75em}
\label{tab:complexity}
\begin{tabular}{c cc cc cc cc} \toprule
\multirow{2}{*}{DType} 
& \multicolumn{2}{c}{Data Transformation} & \multicolumn{2}{c}{Initial Seeding} & \multicolumn{2}{c}{Data Assignment} & \multicolumn{2}{c}{Total}
\\ \cmidrule(lr){2-3} \cmidrule(lr){4-5} \cmidrule(lr){6-7} \cmidrule(lr){8-9}
& Time & Space & Time & Space & Time & Space & Time & Space \\ \midrule
Homo   & $O(dn\log n + n\log^2 n)$ & $O(n\log n)$    & $O(n^{1+\rho} \log^2 n)$ & $O(n^{1+\rho})$ & $O(ndk)$ & $O(n)$ & $O(n^{1+\rho} \log^2 n + ndk)$ & $O(n^{1+\rho})$    \\
Hetero & $O(n^{1+\rho}\log n)$  & $O(n^{1+\rho})$ & $O(n^{1+2\rho}\log n)$   & $O(n^{1+\rho})$ & $O(ndk)$ & $O(n)$ & $O(n^{1+2\rho}\log n + ndk)$   & $O(n^{1+\rho})$ \\
Sparse & $O(dn^{1+\rho}\log n)$ & $O(n^{1+\rho})$ & $O(n^{1+2\rho}\log n)$   & $O(n^{1+\rho})$ & $O(ndk)$ & $O(n)$ & $O(n^{1+2\rho}\log n + ndk)$   & $O(n^{1+\rho})$ \\ \bottomrule
\end{tabular}
\vspace{-0.75em}
\end{table*}

\paragraph{Communication Cost}
In SILK, the majority voting needs the global frequency of data IDs in different buckets to determine the $C_{shared}$'s. However, different processes only have a part of the bins. We need to conduct a synchronization to broadcast bins (including the local bucket IDs and their data IDs) among different processes, but it will lead to a very high communication cost.

In our implementation, we adapt majority voting in SILK to identify $C_{shared}$'s for each process $P_j$ from its local bins. 
After that, each $P_j$ \emph{only} stores $C_{shared}$'s in its local bins. Then, we synchronize the $C_{shared}$'s among different processes such that each $P_j$ has all distinct $C_{shared}$'s together for all bins. 
Since the size of $C_{shared}$ is much smaller than that of the bin itself, the communication cost can be reduced significantly. 
After the synchronization of $C_{shared}$'s, we conduct SILK again for each $P_j$ for deduplication.

\begin{figure}[t]
\centering
\vspace{1.0em}
\includegraphics[width=0.5\textwidth]{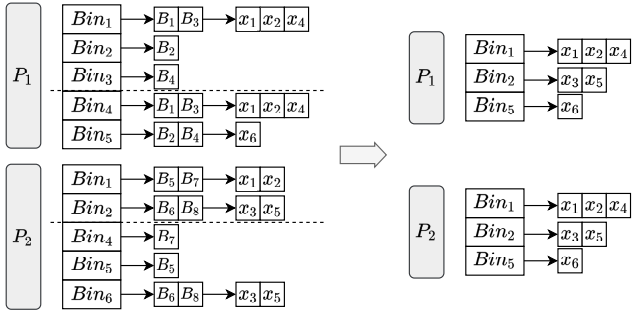}
\vspace{-1.0em}
\caption{An example to reduce communication cost}
\label{fig:communication}
\vspace{-0.5em}
\end{figure}

\begin{example}
\label{exp:communication}
We now use Figure \ref{fig:communication} for an illustration. Suppose the data objects and buckets in Figure \ref{fig:communication} are as same as those in Figure \ref{fig:example}. 
Considering the process $P_1$. For ${Bin}_1$, since the data objects $\bm{x_1},\bm{x_2},\bm{x_4}$ appear more than half time of the buckets $B_1$ and $B_3$, we get $C_{shared}=\{\bm{x_1},\bm{x_2},\bm{x_4}\}$ from ${Bin}_1$. Similarly, we get $C_{shared}=\{\bm{x_1},\bm{x_2},\bm{x_4}\}$ and $C_{shared}=\{\bm{x_6}\}$ from ${Bin}_4$ and ${Bin}_5$, respectively. We ignore ${Bin}_2$ and ${Bin}_3$ because $\num{{Bin}_j} \leq 1$. 
The left part of Figure \ref{fig:communication} shows the $C_{shared}$'s from the local bins of $P_1$ and $P_2$. After that, we conduct SILK again to remove the duplicates $\{\bm{x_1},\bm{x_2},\bm{x_4}\}$ and $\{\bm{x_3},\bm{x_5}\}$ from ${Bin}_4$ and ${Bin}_6$, respectively. The right part of Figure \ref{fig:communication} shows the state after deduplication. 
\hfill $\triangle$ \par 
\end{example}

Since the majority voting is conducted on local bins, the $C_{shared}$'s in Figure \ref{fig:communication} might be slightly different from those in Figure \ref{fig:example}. For example, we get $\{\bm{x_3},\bm{x_5}\}$ instead of $\{\bm{x_3},\bm{x_5},\bm{x_6}\}$ from ${Bin}_2$. However, if the shared data objects are close enough, most of them could be identified from local bins. Thus, we argue that it is reasonable to use this operation to reduce the communication cost by allowing a minor loss inaccuracy.

\subsection{Complexity Analysis} 
\label{sect:impl:complexity}
We now discuss the time and space complexities of GEEK. Let $k$ be the number of central vectors for data assignment, i.e., $k=\num{\mathcal{C}}$. To be concise, we only analyse the time and space complexities for the homogeneous dense data. 

\paragraph{Data Transformation}
We first take $O(mnd)$ time and $O(md)$ space to use $m$ QALSH functions to compute the LSH values. Then, we spend $O(mn\log n)$ time to scan and sort the $m$ hash tables for bucket partitioning. We also use $O(mn)$ space to store buckets (with data IDs only). According to QALSH \cite{huang2015query}, $m = O(\log n)$. Thus, we need $O(dn\log n + n\log^2 n)$ time and $O(n\log n)$ space.

\paragraph{Initial Seeding}
We then analyse the time and space complexities of SILK. 
Suppose $N_\mathcal{B}$ is the number of buckets and $D_\mathcal{B}$ is the average size of data IDs in each bucket. For the homogeneous dense data, $N_\mathcal{B} = mt$ and $D_\mathcal{B}=\frac{n}{t}$, so $N_\mathcal{B} D_\mathcal{B} = mn = O(n\log n)$. Moreover, we can control the granularity of buckets so that $N_\mathcal{B} = O(n)$. 
SILK first uses MinHash \cite{broder1998min} and takes $O(N_\mathcal{B} D_\mathcal{B} KL)$ time and $O(N_\mathcal{B} L)$ space to compute bins. 
The majority voting needs $O(N_\mathcal{B} D_\mathcal{B} L)$ time to scan all buckets and their data IDs in bins to determine $C_{shared}$'s. 
Let $N_\mathcal{C}$ be the total number of data IDs in $\mathcal{C}$, i.e., $N_\mathcal{C} = \sum_{C_{shared} \in \mathcal{C}} \num{C_{shared}}$. We need $O(N_\mathcal{C})$ space to store $\mathcal{C}$. According to MinHash \cite{broder1998min}, $K=O(\log n)$ and $L=O(n^\rho)$, where $\rho = \ln(1/p_1)/\ln(1/p_2)$. Some data IDs may be appeared in multiple $C_{shared}$'s, but $k = \num{\mathcal{C}} \ll n$. Thus, SILK needs $O(N_\mathcal{B} D_\mathcal{B} KL + N_\mathcal{B} D_\mathcal{B} L)=O(n^{1+\rho} \log^2 n)$ time and $O(N_\mathcal{B} L + N_\mathcal{C}) = O(n^{1+\rho})$ space. As can be seen, the time and space complexities of SILK are independent of $k$, and its time complexity is highly dependent on the total number of data IDs among the buckets, i.e., $N_\mathcal{B} D_\mathcal{B}$.

\paragraph{Data Assignment}
In this phase, we first take $O(N_\mathcal{C} d)$ time to compute the $k$ centroids; then, we spend $O(ndk)$ time and $O(n)$ space to achieve the final clusters. As $N_\mathcal{C} = O(n)$, we need $O(ndk)$ time and $O(n)$ space for data assignment. Table \ref{tab:complexity} summarizes the time and space complexities of GEEK among the three types of data.\footnote{DType is short for data type. Homo, Hetero, and Sparse are short for the homogeneous dense data, heterogeneous dense data, and sparse data, respectively.}

\subsection{Discussion}
\label{sect:method:discussion}
Similar to $k$-means and its variants, GEEK is good at discovering compact clusters with a spherical shape. However, it may fail to find complex-shaped clusters like DBSCAN \cite{ester1996density, schubert2017dbscan}, or form dendrogram to facilitate the understanding of these clusters like hierarchical methods such as BIRCH \cite{zhang1996birch} and CURE \cite{guha1998cure}.

Nevertheless, as discussed in Section \ref{sect:impl:complexity}, the time complexity of SILK is independent of $k$. With this characteristic, we can first use GEEK to form a large number of microclusters and then use density-based methods to build on top of it to identify complex-shaped clusters. Moreover, we can build an index for the microclusters to speed up the NNS (e.g., FAISS \cite{johnson2019billion}) and improve the scalability of other kinds of clustering methods such as DBSCAN and BIRCH. Thus, besides discovering the clusters, GEEK can be a fundamental tool to support and accelerate other clustering methods.

%

\section{Experiments}
\label{sect:expt}
We study the performance of GEEK over five large-scale datasets. All experiments are conducted on a distributed CPU-GPU platform which consists of 2 computing nodes and 6 GPUs. 
Node 1 has an Intel(R) Xeon(R) Platinum 8170 CPU @ 2.10 GHz with 52 cores, 1 TB RAM, and 2 Nvidia GeForce GTX 1080 Ti's with 11 GB memory. 
Node 2 has an Intel(R) Xeon(R) Gold 6130 CPU @ 2.10 GHz with 32 cores, 256 GB RAM, and 4 Nvidia GeForce GTX 1080 Ti's with 11 GB memory. Both machines are running on CentOS 7.4.

\subsection{Experiment Setup}
\label{sect:expt:setup}
\vspace{-0.5em}
\paragraph{Competitors}
To the best of our knowledge, GEEK is the first generic framework that supports clustering over different types of data, so there is no direct competitor. Alternatively, we select different state-of-the-art GPU-based methods as its competitors for different types of data. Specifically, for the homogeneous dense data, we choose Lloyd \cite{lloyd1982least}, Yinyang \cite{ding2015yinyang},\footnote{Lloyd and Yinyang are available at \url{https://github.com/src-d/kmcuda}.} and FAISS \cite{johnson2019billion}\footnote{FAISS is a very popular GPU-based library for efficient similarity search and clustering, its implementation is available at \url{https://github.com/facebookresearch/faiss}.} as baselines; for the heterogeneous dense data and the sparse data, we implement a GPU-based $k$-modes \cite{huang1998extensions} as its competitor.

\paragraph{Datasets}
We evaluate the performance of GEEK over five real-life datasets, i.e., Gist, Sift10M, Sift1B,\footnote{Gist, Sift10M, and Sift1B can be downloaded from \url{http://corpus-texmex.irisa.fr/}.} GeoNames,\footnote{\url{https://www.kaggle.com/geonames/geonames-database}.} and URL.\footnote{\url{https://archive.ics.uci.edu/ml/datasets/URL+Reputation}.} 
For the homogeneous dense datasets Gist, Sift10M, and Sift1B, we use Euclidean distance, while for the heterogeneous dense dataset GeoNames and the sparse dataset URL, we use the distance $(1-\text{Jaccard})$. The statistics of datasets are summarized in Table \ref{tab:data}. Their details are described as follows.
\begin{itemize}
\item \textbf{Gist.} This homogeneous dense dataset has $10^6$ Gist image features, where each feature is a 960-dimensional vector.

\item \textbf{Sift10M, Sift1B.} These two homogeneous dense datasets consist of $10^7$ and $10^9$ Sift features from images, respectively, where each feature is a 128-dimensional vector. 

\item \textbf{GeoNames.} It contains over 11 million geographical names. After data cleaning, we get a heterogeneous dense dataset with 9 attributes, i.e., latitude, longitude, feature class, feature code, country code, population, elevation, dem, timezone.

\item \textbf{URL.} This sparse dataset collects 2.3 million URLs, where it has around 3.2 million features but with only around 116 non-zeros for each data object \cite{ma2009identifying}. We adopt DOPH \cite{shrivastava2014densifying} to reduce the dimensionality to $400$.
\end{itemize}

\paragraph{Evaluation Metrics}
As done in FAISS \cite{johnson2019billion}, one of the most frequent applications for clustering is to use the clusters for large-scale NNS. For this case, we do not know the ground truth of clusters. The compactness of clusters becomes a vital indicator to evaluate the effectiveness of a clustering method. We use the following evaluation metrics for performance evaluation.
\begin{itemize}
\item \textbf{Time.} We use the running time (or simply time) to evaluate the efficiency of a method. It is defined as the wall-clock time (in seconds) of a method for clustering.

\item \textbf{Radius.} We use radius to evaluate the effectiveness of a method. Given a cluster of $m$ data objects $\{\bm{x_1},\cdots,\bm{x_m}\}$ and the central vector $\bm{c}$, $radius = \max_{i \in \{1,\cdots,m\}} Dist(\bm{x_i},\bm{c})$. A small radius indicates a compact cluster.
\end{itemize} 

We report the mean results of these metrics over five times of repetition. The number of clusters in FAISS \cite{johnson2019billion} is fixed because it splits the large clusters into two when $k$ reduces. However, Lloyd \cite{lloyd1982least}, Yinyang \cite{ding2015yinyang}, $k$-modes \cite{huang1998extensions}, and GEEK do not implement this operation. Their final number of clusters (i.e., $k^*$) may be smaller than the pre-specified $k$ value. Since the radius is highly sensitive to $k^*$, we focus on $k^*$ during the performance evaluation. 

\begin{table}[t]
\caption{Statistics of five real-life datasets}
\vspace{-0.5em}
\label{tab:data}
\begin{tabular}{cccccc} \toprule
Datasets & $n$  & $d$  & DType & Distance \\ \midrule
Gist     & $1.0 \times 10^6$ & $960$ & Homo   & Euclidean  \\
Sift10M  & $1.0 \times 10^7$ & $128$ & Homo   & Euclidean \\
Sift1B   & $1.0 \times 10^9$ & $128$ & Homo   & Euclidean \\
GeoNames & $1.1 \times 10^7$ & $9$   & Hetero & $1-\text{Jaccard}$ \\
URL      & $2.3 \times 10^6$ & $3.2 \times 10^6$ & Sparse & $1-\text{Jaccard}$ \\ \bottomrule 
\end{tabular}
\vspace{-0.5em}
\end{table}

\begin{figure*}[htb]
\centering
\includegraphics[width=1.0\textwidth]{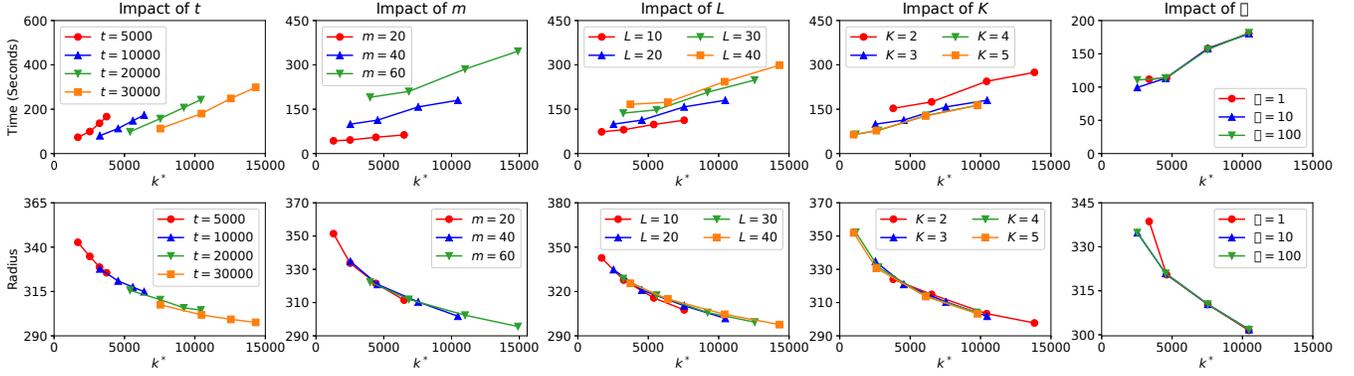}
\vspace{-1.5em}
\caption{The results of GEEK over Sift10M under different parameter settings}
\label{fig:param}
\end{figure*}

\begin{figure*}[htb]
\centering
\includegraphics[width=1.0\textwidth]{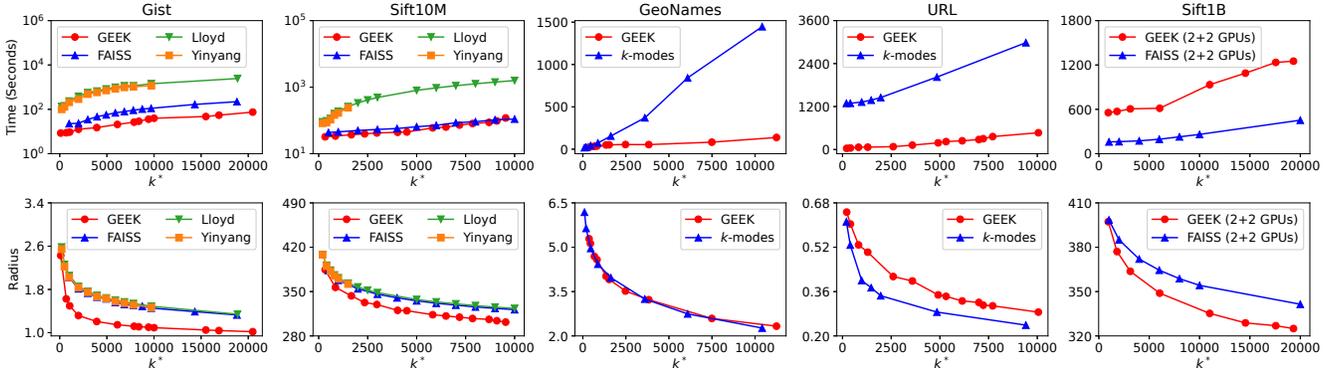}
\vspace{-1.5em}
\caption{The results of all methods}
\label{fig:results}
\end{figure*}

\subsection{Parameters Studies}
\label{sect:expt:para}
We first study the impacts of parameters of GEEK in a single GPU fashion. To be concise, we only show the results of GEEK over Sift10M, where $\mathcal{B}$ is determined by $t$ and $m$. Thus, we study the impacts of $t$, $m$, $L$, $K$, and $\delta$. The results are depicted in Figure \ref{fig:param}.

\vspace{-0.1em}
\paragraph{Impact of $t$}
We fix $m=40$, $K=3$, $\delta=10$ and vary $t \in \{5000,10000$, $20000,30000\}$. We use different $L \in \{10,20,30,40\}$ to get different $k^*$'s. 
From Figure \ref{fig:param}, GEEK enjoys its best performance under different $t$'s for different ranges of $k^*$. Since $t$ controls the granularity of buckets, a larger $t$ leads to a better performance for a larger $k^*$. 
Thus, we use different $t \in \{5000,10000,20000,30000\}$ to get different $k^*$ values for the remaining parameters studies.

\vspace{-0.1em}
\paragraph{Impacts of $m$ and $L$}
We first fix $L=20$, $K=3$, $\delta=10$ and vary $m \in \{20,40,60\}$. For different ranges of $k^*$, GEEK achieves its best performance under different $m$'s. For example, when $k^* > 11,000$, GEEK enjoys the least running time and comparable radius values under $m=60$, while for $k^*<5,000$, GEEK achieves its best results under $m=20$.
We then fix $m=40$, $K=3$, $\delta=10$ and vary $L \in \{10,20,30,40\}$. Similar to the impacts of $t$ and $m$, GEEK achieves its best performance under different $L$'s. 
As the increase of $m$ or $L$, more buckets will be generated for seeding and results in a larger $k^*$.
Based on the above observations, we tune a combination of $t$, $m$, and $L$ to get the best result of GEEK for different ranges of $k^*$.

\vspace{-0.1em}
\paragraph{Impacts of $K$ and $\delta$}
We first fix $m=40$, $L=20$, $\delta=10$ and vary $K \in \{2,3,4,5\}$. Except $K=2$, the results of GEEK under other settings of $K$ are comparable to each other. Thus, the performance of GEEK is less sensitive to $K$. 
Then, we fix $m=40$, $L=20$, $K=3$ and vary $\delta \in \{1,10,100\}$. Similar to the impact of $K$, the results of GEEK under different $\delta$'s are almost identical to each other. 
For simplicity, we set $K=3$ and $\delta = 10$ by default.

\vspace{-0.1em}
\paragraph{Summary}
In the subsequent experiments, we set $K=3$ and $\delta=10$ for GEEK; to remove the impact of parameters, we run GEEK on a grid search of $t \in \{5000,10000,20000,30000\}$, $m \in \{20,40,60\}$, and $L \in \{10,20,30,40\}$ and report its best results for different $k^*$ values. For Yinyang and FAISS, we use the settings suggested by the authors \cite{ding2015yinyang,johnson2019billion} and set $k$ from $100$ to $20,000$ to get different $k^*$'s if possible. Lloyd and $k$-modes do not require extra parameters. 

\subsection{Results and Analysis}
\label{sect:expt:results}
We evaluate GEEK in terms of five aspects: the clustering performance, the scalability to massive datasets, the initial seeding performance and the performance on multi-GPUs and multi-nodes. 

\paragraph{Clustering Performance}
We first study the performance of GEEK under different $k^*$ values. For the competitors, we report their least running time as soon as they converge. Since Yinyang only supports a single GPU, we run all methods on 1 GPU to make a fair comparison. The results are shown in Figure \ref{fig:results}.

GEEK is slightly faster than FAISS over Gist and Sift10M, and compared with Lloyd and Yinyang, GEEK achieves at least one order of magnitude acceleration when $k^* > 2,000$. 
The reason is that the phases of data transformation and initial seeding of GEEK are independent of $k$, and it conducts a one-pass data assignment only. Thus, GEEK becomes increasingly efficient as $k$ increases. 
Moreover, GEEK enjoys a smaller radius than the competitors so that it can generate more compact clusters than the competitors. Therefore, GEEK outperforms Lloyd, Yinyang, and FAISS over these homogeneous dense datasets.

As can be seen from Figure \ref{fig:results}, GEEK also enjoys at least one order of magnitude less running time than $k$-modes over GeoNames and URL when $k^* > 5,000$. 
The radius values of GEEK are close to (or slightly larger than) those of $k$-modes, which indicates that GEEK can generate clusters as compact as $k$-modes. 
Notice that GEEK is much more efficient than $k$-modes, and as will be shown later, GEEK can support multi-GPUs and multi-nodes. 
If the compactness is much more important than the efficiency, we can conduct several passes of data assignments for GEEK with multi-GPUs to get more compact clusters yet without losing efficiency. Thus, we can claim that GEEK outperforms $k$-modes over the heterogeneous dense dataset GeoNames and the sparse dataset URL. 

Additionally, as $k^*$ increases, the slope of GEEK is smooth like FAISS, which is much smaller than those of Lloyd, Yinyang, and $k$-modes. This observation further demonstrates that the running time of GEEK is not very sensitive to $k^*$. Due to the large GPU memory usage, Yinyang cannot support clustering with large $k^*$ values. Thus, the curves of Yinyang are relatively short.

\paragraph{Scalability to Massive Datasets}
We then study the scalability of GEEK on a billion-scale dataset Sift1B. Since the performance of Lloyd and Yinyang are much worse than GEEK and FAISS, we only run GEEK and FAISS on a distributed CPU-GPU platform with 2 computing nodes, where each computing node uses 2 GPUs (2+2 GPUs). The results are displayed in Figure \ref{fig:results}.

The radius values of GEEK are much smaller than those of GEEK, especially when $k^* > 5,000$. However, FAISS is $2 \sim 4$ times faster than GEEK because FAISS uses a sampling method in its implementation to reduce the computational cost and space overhead. 
Given a specific $k$ value, FAISS uniformly draws a sample set of $(256 \cdot k)$ data to determine $k$ centroids. The sample size is much smaller than $10^9$, e.g., only $0.5\%$ of Sift1B for a considerable $k$ value such as $k=20,000$. Thus, FAISS enjoys better scalability than GEEK. However, using a sample set to deal with billion-scale data might be problematic. The sample set is merely a sketch representation of the full data. The centroids may concentrate on the high-density areas, leading to a large radius for clusters with data objects from low-density areas.
In fact, the scalability of GEEK is still good enough for most applications. According to Figure \ref{fig:results}, GEEK can provide compact clusters for $k^* \approx 20,000$ with less than 20 minutes.

\begin{figure}[t]
\centering
\includegraphics[width=0.48\textwidth]{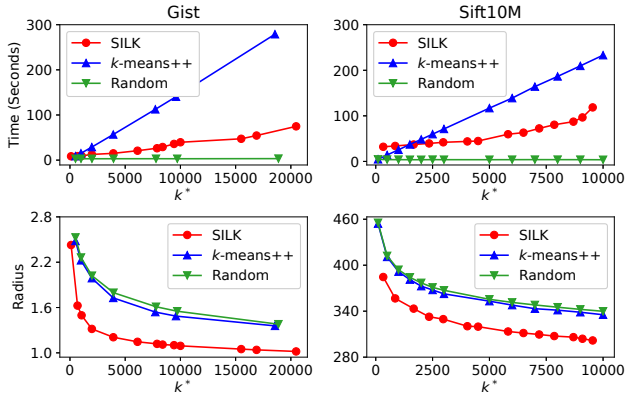}
\vspace{-1.5em}
\caption{The results of different initial seeding methods over Gist and Sift10M}
\label{fig:seeding}
\end{figure}

\paragraph{Initial Seeding Performance}
Next, we study the performance of SILK. We choose $k$-means++ \cite{arthur2007k} and the random selection (Random)\footnote{The GPU versions of $k$-means++ and Random (implemented by CUDA) are available at \url{https://github.com/src-d/kmcuda}.} as competitors. 
To make a fair comparison, we only record their running time of initial seeding; then, we use their seeds for a one-pass data assignment to evaluate the seeding performance. 
We show the results on 1 GPU over Gist and Sift10M in Figure \ref{fig:seeding}. Similar trends can be observed from other datasets.

The radius values of SILK are much smaller than those of $k$-means++ and Random, which indicates that the initial seeds generated by SILK are \emph{well-selected} and can be used to achieve more \emph{compact} clusters than others. 
Even though Random enjoys the least running time, SILK requires less running time than $k$-means++, and its advantage is more apparent when $k^*$ increases. 
Moreover, the running time of $k$-means++ is linear to $k^*$, while that of SILK and Random is less sensitive to $k^*$. 
This phenomenon could be because the time complexity of both SILK and Random is independent of $k$, while $k$-means++ requires $O(k)$ iterations to generate the same number of seeds.
Thus, SILK outperforms $k$-means++ and Random over Gist and Sift10M, especially in large $k^*$ values.

\begin{figure}[t]
\centering
\includegraphics[width=0.48\textwidth]{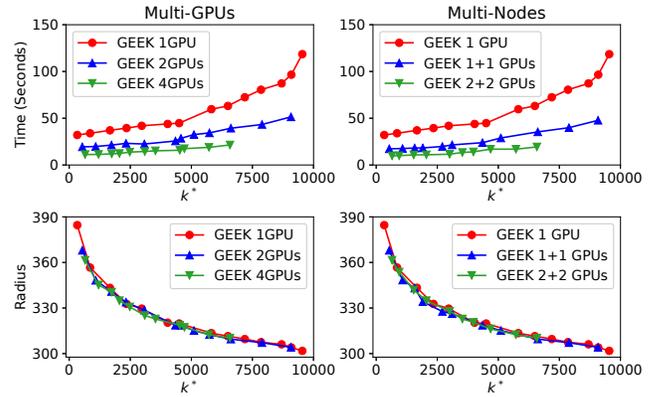}
\vspace{-1.5em}
\caption{The results of GEEK over Sift10M on multi-GPUs and multi-nodes}
\label{fig:multi}
\end{figure}

\paragraph{Performance on Multi-GPUs}
To verify the effectiveness of the distributed implementation, we first study the performance of GEEK on multi-GPUs. 
To reduce the impact of communication overhead from different computing nodes, we run GEEK on a single node (i.e., Node 2) and vary the number of GPUs in a range of $\{1,2,4\}$. 
The results of GEEK over Sift10M are displayed in Figure \ref{fig:multi}. Similar trends can be observed from other datasets.

The acceleration of GEEK is asymptotically linear to the number of GPUs, which indicates that each GPU's workload for the original data and intermediate data is \emph{balanced}. 
Besides, the radius values of GEEK under different number of GPUs are close to each other. Thus, GEEK can provide stable, compact results on multi-GPUs.

\paragraph{Performance on Multi-Nodes}
To this end, we study the performance of GEEK on multi-nodes. 
We consider three cases: (1) Node 1 uses one GPU only (1 GPU); (2) both Node 1 and Node 2 use one GPU (1+1 GPUs); (3) both Node 1 and Node 2 use two GPUs (2+2 GPUs). 
The results over Sift10M are depicted in Figure \ref{fig:multi}. Similar trends can be observed from other datasets.

Similar to the results on multi-GPUs, the acceleration of GEEK is asymptotically linear to the total number of GPUs, demonstrating the effectiveness of the distributed implementation in reducing the \emph{communication cost} across different nodes. 
Since the radius values under different cases are close to each other, GEEK can provide stable, compact clusters on multi-nodes. 
Thus, no matter on a single node with multi-GPUs or multi-nodes with multi-GPUs, the distributed implementation of GEEK is efficient and effective.


\section{Related work}
\label{sect:related}
Clustering is a fundamental problem with extensive studies. 
Well known methods includes partitioning \cite{macqueen1967some, lloyd1982least, lucasius1993k, huang1998extensions, ng2002clarans}, hierarchical \cite{zhang1996birch, guha1998cure}, density-based \cite{ester1996density, ankerst1999optics, gan2015dbscan, schubert2017dbscan} and grid-based methods \cite{wang1997sting, agrawal1998automatic, sheikholeslami1998wavecluster}. A comprehensive survey can be referred to \cite{han2001spatial, xu2015comprehensive}. 
Since GEEK is most related to partitioning methods such as $k$-means and its variants, we briefly discuss those methods.

$k$-means was first introduced by MaxQueen \cite{macqueen1967some}. 
The most famous version is \emph{Lloyd's algorithm} \cite{lloyd1982least}, which randomly selects $k$ data objects as initial seeds and conducts the assign-update iteration with $O(ndk)$ time until the clusters converge. As data have been rapidly increased, many variants have been proposed to improve Lloyd's algorithm in initial seed selection, the support of large $d$ and $k$, and the acceleration of massive data. 

As is well-known, a good selection of initial seeds is crucial for $k$-means to quickly converge. The classic remedy is to use $k$-means++\cite{arthur2007k}, but it requires $O(ndk^2)$ time and $O(k)$ iterations, which is hard to parallel. Bahmani et al. \cite{bahmani2012scalable} proposed a parallel variant $k$-means$\parallel$ and reduced the iterations to $O(\log k)$. To get a fast and provably good seeding algorithm, Bachem et al. \cite{bachem2016fast} designed a Markov chain Monte Carlo sampling method to generate $k$ seeds that satisfy the $k$-means++ distribution. Later, they provided a new analysis of $k$-means$\parallel$ that bounds the quality of $k$ seeds for any number of iterations \cite{bachem2017distributed}. However, these methods still have a running time of $\Omega(k^2)$. Recently, Cohen-addad et al. proposed a near linear time method for $k$-means++ seeding with rejection sampling. Compared with existing techniques, SILK is independent of $k$ and it does not need to set up $k$ in advance. Thus, GEEK can efficiently generate a large number of initial seeds. 

Another direction to optimize $k$-means is to support large values of $d$ and $k$. Elkan \cite{elkan2003using} introduced a representative work which uses triangle inequality to avoid unnecessary distance computations. Algorithms of the same kinds include Hamerly \cite{hamerly2010making}, Ding et al. \cite{ding2015yinyang}, Newling et al. \cite{newling2016fast, newling2016nested}, and Curtin \cite{curtin2017dual}. Since we aim to support different types of data, we cannot leverage the geometric properties such as triangle inequality for acceleration. Instead, we apply GPUs and implement GEEK on a distributed CPU-GPU platform to speed up the distance computation for data assignment.

Recently, many researchers also consider using modern hardwares, such as GPUs \cite{cao2006scalable, li2013speeding, kohlhoff2013k} and many-core supercomputers \cite{hadian2014high, li2018large}, to scale up $k$-means for massive datasets. 
However, they are highly optimized for a specific distance, e.g., Euclidean distance. In contrast, GEEK is a generic distributed clustering framework, which can handle large-scale data with many distance functions.

\section{Conclusion}
\label{sect:conclusion}
In this paper, we introduce a novel generic distributed clustering framework GEEK to process massive data of three data types. We also propose a new seeding method SILK to efficiently generate seeds from similar buckets. As distinct from $k$-means++ and its variants, SILK does not need to pre-specify $k$, and its time complexity is independent of $k$. With this advantage, GEEK can also be a fundamental tool to support and accelerate other clustering methods. We design and implement GEEK on a distributed CPU-GPU platform. 
Extensive experiments over five large-scale datasets demonstrate the superior performance of GEEK, especially when $k$ is large. 


\balance

\bibliographystyle{ACM-Reference-Format}
\bibliography{FullPaper}

\end{document}